\documentclass[11pt,a4paper]{article}

\usepackage{amssymb}
\usepackage{graphicx}
\usepackage{bm}

\usepackage{mathrsfs}
\usepackage{cite}

\newcommand*\xbar[1]{%
  \hbox{%
    \vbox{%
      \hrule height 0.5pt 
      \kern0.5ex
      \hbox{%
        \kern-0.1em
        \ensuremath{#1}%
        \kern-0.1em
      }%
    }%
  }%
}

\begin{document}

\title{All-loop renormalization group invariants for MSSM}

\author{
D.M.Rystsov and K.V.Stepanyantz\\
\\
{\small{\em Moscow State University}}, {\small{\em  Faculty of Physics, Department  of Theoretical Physics}}\\
{\small{\em 119991, Moscow, Russia}}\\
}

\maketitle

\begin{abstract}
For MSSM from the gauge couplings, Yukawa couplings, and the coefficient $\bm{\mu}$ in the part of superpotential quadratic in the Higgs superfields we construct combinations which (for certain renormalization prescriptions) do not depend on the renormalization point in all loops. In other words, these combinations are the renormalization group invariants. Similar invariants are also constructed for NMSSM. The derivation is based on the nonrenormalization of the superpotential and the NSVZ equations. We argue that the scale invariance of the considered combinations takes place in the class of the HD+MSL schemes. This fact has been verified in the lowest orders, up to and including the one in which the dependence on the renormalization prescription becomes essential. It is also demonstrated that in the $\overline{\mbox{DR}}$ scheme the renormalization group invariance does not take place starting from the approximation, where the scheme dependence manifests itself.
\end{abstract}

\section{Introduction}
\hspace*{\parindent}

The detailed analysis of quantum corrections can tell a lot about the world around us. For instance, the renormalization group running of gauge couplings in the supersymmetric extensions of the Standard model \cite{Ellis:1990wk,Amaldi:1991cn,Langacker:1991an} is a strong indication in favor of supersymmetry \cite{West:1990tg,Buchbinder:1998qv} and Grand Unification \cite{Mohapatra:1986uf}. Another interesting prediction of the Grand Unified Theories is the unification of the Yukawa couplings \cite{Chanowitz:1977ye,Buras:1977yy,Georgi:1979ga}, which leads to certain relations between masses of elementary particles. However, it is very difficult to match such predictions with the experimental data \cite{ParticleDataGroup:2024cfk} because a theory describing physics beyond the Standard model is not exactly known. At present, one of the most promising candidates for this is supersymmetric extensions of the Standard model. The simplest of them is the Minimal Supersymmetric Standard Model (MSSM), which is a theory invariant under the $SU(3)\times SU(2)\times U(1)$ gauge transformations with softly broken supersymmetry, see, e.g., \cite{Mohapatra:1986uf}. However, the soft terms do not affect the renormalization of the rigid theory and, in particular, such parameters as the gauge and Yukawa couplings. That is why a lot of MSSM quantum properties can be investigated using the results obtained for ${\cal N}=1$ supersymmetric theories. For instance, the renormalization group running of the gauge couplings can be described with the help of NSVZ $\beta$-functions \cite{Novikov:1983uc,Jones:1983ip,Novikov:1985rd,Shifman:1986zi} in all orders, which for MSSM were first written in \cite{Shifman:1996iy}. Various phenomenological applications of these exact relations can be found in, e.g., \cite{Ghilencea:1997mu,Amelino-Camelia:1998xgn,Mihaila:2013wma,Heinemeyer:2019vbc}.

The existence of the exact equations relating the $\beta$-functions to the anomalous dimensions of the chiral matter superfields can lead to some interesting consequences. For instance, according to
\cite{INR_TH}, in ${\cal N}=1$ SQCD interacting with the Abelian gauge superfield in a supersymmetric way the running of the strong gauge coupling $\alpha_s$ is related to the running of the electromagnetic gauge coupling $\alpha$. Assuming that there are $N_f$ identical flavors consisting of the chiral superfields $\phi_{\mbox{\scriptsize a}}$ and $\widetilde\phi_{\mbox{\scriptsize a}}$ in the irreducible representations $R$ and $\xbar{R}$ of the non-Abelian gauge group $G$ with opposite $U(1)$ charges, this relation can be presented in the form of the renormalization group invariant (RGI)

\begin{equation}\label{SQCD+SQED_RGI}
\Big(\frac{\alpha_s}{\mu^3}\Big)^{C_2} \exp\Big(\frac{2\pi}{\alpha_s} - \frac{T(R)}{\mbox{dim}\,R}\cdot \frac{2\pi}{\alpha}\Big) = \mbox{RGI}.
\end{equation}

\noindent
Here $\mbox{tr}(T^A T^B) = T(R)\delta^{AB}$, where $T^A$ are the generators in the representation $R$ (of the dimension $\mbox{dim}\,R$). The Casimir $C_2$ is defined by the equation $f^{ACD} f^{BCD} = C_2 \delta^{AB}$, where $f^{ABC}$ are the structure constants.

Eq. (\ref{SQCD+SQED_RGI}) is obtained by excluding the anomalous dimension of the matter superfields from the NSVZ equations.\footnote{If the flavors have different electric charges, then it is possible to relate the $\beta$-function of ${\cal N}=1$ SQCD to the Adler $D$-function \cite{Adler:1974gd} using the NSVZ-like equation for it \cite{Shifman:2014cya,Shifman:2015doa}.} Therefore, exactly as the NSVZ relations, it is satisfied only for certain renormalization prescriptions and should be valid in an arbitrary NSVZ scheme. Such schemes can naturally be constructed with the help of the higher covariant derivative regularization \cite{Slavnov:1971aw,Slavnov:1972sq,Slavnov:1977zf} formulated in ${\cal N}=1$ superspace \cite{Krivoshchekov:1978xg,West:1985jx}. According to \cite{Kataev:2013eta,Stepanyantz:2020uke}, the NSVZ equations are valid for the renormalization group functions (RGFs) defined in terms of the bare couplings for an arbitrary renormalization prescription supplementing the higher covariant derivative regularization and for (standard) RGFs defined in terms of the renormalized couplings in the HD+MSL scheme.\footnote{This statement has been verified by some explicit multiloop calculations in such orders of the perturbation theory where the scheme dependence becomes essential \cite{Kazantsev:2018nbl,Kuzmichev:2019ywn,Aleshin:2020gec,Shirokov:2023jya}.} In the latter case for a theory regularized by higher covariant derivatives the renormalization is made with the help of minimal subtractions of logarithms  \cite{Kataev:2013eta,Shakhmanov:2017wji}. In the $\overline{\mbox{DR}}$ scheme (when a theory is regularized by dimensional reduction \cite{Siegel:1979wq} and divergences are removed by modified minimal subtraction \cite{Bardeen:1978yd}) Eq. (\ref{SQCD+SQED_RGI}) is not valid \cite{INR_TH} exactly as the NSVZ equation(s) \cite{Jack:1996vg,Jack:1996cn,Jack:1998uj,Harlander:2006xq}.

However, it would be interesting to consider a more realistic theory like MSSM, which is the simplest supersymmetric extension of the Standard model. Although in this model supersymmetry is softly broken, the renormalization of the parameters entering the rigid part of its action occurs exactly as for the supersymmetric theories. The renormalization of the soft terms can be related to the renormalization of the rigid theory \cite{Hisano:1997ua,Jack:1997pa,Avdeev:1997vx}. For example, the running of the gaugino masses is determined by all-loop RGIs constructed in \cite{Hisano:1997ua}.

In this paper for MSSM we construct two independent RGIs analogous to (\ref{SQCD+SQED_RGI}). For this purpose the anomalous dimensions of the chiral matter superfields are excluded from the all-loop equations describing the renormalization group running of the gauge couplings, Yukawa couplings and the $\bm{\mu}$ parameter. One of the resulting expressions does not contain the renormalization point $\mu$. Another similar expression explicitly depends on $\mu$, but is independent of the $\bm{\mu}$ parameter in the MSSM superpotential.

The paper is organized as follows. In Sect. \ref{Section_Renormalization} we recall some basic information about MSSM and write down the all-loop exact equations describing the renormalization group evolution of some its parameters. In Sect. \ref{Section_RGI_MSSM} we exclude the anomalous dimensions of the chiral matter superfields from these equations and obtain two relations restricting the running of the gauge and Yukawa couplings in all orders in the form of RGIs analogous to (\ref{SQCD+SQED_RGI}). Modifications of the results to the case of NMMSM are discussed in Sect. \ref{Section_NMSSM}. The renormalization group invariance of the constructed expressions is verified by the explicit three-loop calculation in Sect. \ref{Section_Verification}, where we demonstrate that it is really valid in the HD+MSL scheme, but is not valid in the $\overline{\mbox{DR}}$ one. The results are briefly discussed in Conclusion, and Appendix \ref{Appendix_Details} contains technical details of the calculations. The explicit expressions for RGFs in the HD+MSL and $\overline{\mbox{DR}}$ schemes are presented in Appendices \ref{Appendix_RGFs_HD+MSL} and \ref{Appendix_RGFs_DR}, respectively.

\section{Exact equations describing the renormalization of the MSSM parameters}
\hspace*{\parindent}\label{Section_Renormalization}

MSSM is an extension of the Standard model, which is a gauge theory with softly broken supersymmetry and the gauge group

\begin{equation}
G = SU(3) \times SU(2) \times U(1)_Y.
\end{equation}

\noindent
This implies that it contains three gauge couplings

\begin{equation}\label{MSSM_Couplings}
\alpha_3 = \frac{e_3^2}{4\pi};\qquad \alpha_2 = \frac{e_2^2}{4\pi};\qquad  \alpha_1 = \frac{5}{3}\cdot \frac{e_1^2}{4\pi}
\end{equation}

\noindent
corresponding to the subgroups $SU(3)$, $SU(2)$, and $U(1)$, respectively. (The factor $5/3$ in the coupling constant $\alpha_1$ is introduced in order that the unification of couplings has the form $\alpha_1=\alpha_2=\alpha_3$.) It is convenient to formulate the theory with the help of ${\cal N}=1$ superfields using the spurion technique \cite{Girardello:1981wz} for the soft terms. In this case quarks, leptons and Higgs doublets are components of the chiral superfields listed in Table \ref{Table_MSSM_Superfields_List}, where, in particular, the superfields including the (charge conjugated) left quarks and leptons can be written in the form

\begin{equation}
Q_I = \left(\begin{array}{c}\widetilde U\\ \widetilde D\end{array}\right)_I;\qquad L_I = \left(\begin{array}{c}\widetilde N\\ \widetilde E\end{array}\right)_I.
\end{equation}

\noindent
In Table \ref{Table_MSSM_Superfields_List} we also present quantum numbers of all chiral matter superfields with respect to the gauge group (representations for $SU(3)$ and $SU(2)$, and hypercharges for $U(1)$). In our notation, the indices $I$, $J$ range from 1 to 3 and numerate generations.

\medskip

\begin{table}[h]
\begin{center}
\begin{tabular}{|c||c|c|c|c|c||c|c|}
\hline
$\mbox{Superfield}\vphantom{\Big(}$ & $Q_I$ & $U_I$ & $D_I$ & $L_I$ &\, $E_I$\, &\ $H_u$\ &\ $H_d$\ \  \\
\hline
\hline
$SU(3)\vphantom{\Big(}$ & $\bar 3$ & 3 & 3 & 1 & 1 & 1 & 1 \\
\hline
$SU(2)\vphantom{\Big(}$ & 2 & 1 & 1 & 2 & 1 & 2 & 2 \\
\hline
$U(1)_Y\vphantom{\Big(}$ & $-1/6$ & $2/3$ & $-1/3$ & $1/2$ & $-1$ & $-1/2$ & $1/2$\\
\hline
\end{tabular}
\end{center}
\caption{MSSM chiral matter superfields and their quantum numbers with respect to the various parts of the gauge group.}\label{Table_MSSM_Superfields_List}
\end{table}

The MSSM action also contains dimensionless Yukawa couplings $(Y_U)_{IJ}$, $(Y_D)_{IJ}$, and $(Y_E)_{IJ}$ (which are $3\times 3$ matrices) inside

\begin{equation}
\Delta S = \frac{1}{2} \int d^4x\, d^2\theta\, W + \mbox{c.c.},
\end{equation}

\noindent
where the superpotential is given by the expression

\begin{eqnarray}\label{MSSM_Superpotential}
&&\hspace*{-9mm} W = \left(Y_U\right)_{IJ}
\left(\widetilde U\ \widetilde D \right)^{a}_I
\left(
\begin{array}{cc}
0 & 1\\
-1 & 0
\end{array}
\right)
\left(
\begin{array}{c}
H_{u1}\\ H_{u2}
\end{array}
\right) U_{aJ}
+ \left(Y_D\right)_{IJ}
\left(\widetilde U\ \widetilde D \right)^{a}_I
\left(
\begin{array}{cc}
0 & 1\\
-1 & 0
\end{array}
\right)
\left(
\begin{array}{c}
H_{d1}\\ H_{d2}
\end{array}
\right)
\nonumber\\
&&\hspace*{-9mm} \times D_{aJ} + \left(Y_E\right)_{IJ} \left(\widetilde N\ \widetilde E \right)_{I}
\left(
\begin{array}{cc}
0 & 1\\
-1 & 0
\end{array}
\right)
\left(
\begin{array}{c}
H_{d1}\\ H_{d2}
\end{array}
\right) E_J
+ \bm{\mu} \left(H_{u1}\ H_{u2} \right)
\left(
\begin{array}{cc}
0 & 1\\
-1 & 0
\end{array}
\right)
\left(
\begin{array}{c}
H_{d1}\\ H_{d2}
\end{array}
\right).
\end{eqnarray}

\noindent
Moreover, the superpotential contains a term with the parameter $\bm{\mu}$, which has the dimension of mass. The color index $a$ in Eq. (\ref{MSSM_Superpotential}) ranges from 1 to 3.

The renormalization group running of the gauge couplings in MSSM can be described exactly in all loops with the help of the NSVZ $\beta$-functions \cite{Shifman:1996iy}. They can be written in the form \cite{Korneev:2021zdz}

\begin{eqnarray}\label{MSSM_Beta_NSVZ}
&&\hspace*{-5mm} \frac{\beta_1}{\alpha_1^2} = - \frac{3}{5} \cdot \frac{1}{2\pi}\bigg[-11 + \mbox{tr}\Big(\frac{1}{6} \gamma_{Q} + \frac{4}{3} \gamma_{U} + \frac{1}{3} \gamma_{D} + \frac{1}{2} \gamma_{L}
+ \gamma_{E}\Big) + \frac{1}{2} \gamma_{H_u} + \frac{1}{2} \gamma_{H_d}\bigg];\quad\nonumber\\
&&\hspace*{-5mm} \frac{\beta_2}{\alpha_2^2} = - \frac{1}{2\pi(1 - \alpha_2/\pi)} \bigg[-1 + \mbox{tr}\Big(\frac{3}{2} \gamma_{Q} + \frac{1}{2} \gamma_{L}\Big) + \frac{1}{2} \gamma_{H_u} + \frac{1}{2} \gamma_{H_d}\bigg];\nonumber\\
&&\hspace*{-5mm} \frac{\beta_3}{\alpha_3^2} = - \frac{1}{2\pi(1 - 3\alpha_3/2\pi)} \bigg[3 + \mbox{tr} \Big(\gamma_{Q} + \frac{1}{2} \gamma_{U} + \frac{1}{2} \gamma_{D}\Big)\bigg]
\end{eqnarray}

\noindent
and relate three gauge $\beta$-functions of the theory to the anomalous dimensions of the chiral matter superfields. They are defined by the equations

\begin{equation}
\beta_i(\alpha,Y) = \frac{d\alpha_i}{d\ln\mu}\bigg|_{\alpha_0,Y_0=\mbox{\scriptsize const}};\qquad \gamma_i(\alpha,Y) = \frac{d\ln Z_i}{d\ln\mu}\bigg|_{\alpha_0,Y_0=\mbox{\scriptsize const}},
\end{equation}

\noindent
where the subscript $0$ denotes the bare values.

RGFs describing the renormalization of the Yukawa couplings and of the parameter $\bm{\mu}$ can also be related to the anomalous dimensions of the matter superfields due to the nonrenormalization of the superpotential \cite{Grisaru:1979wc}. Taking into account that the superpotential does not receive divergent quantum corrections we see that there are renormalization prescriptions for which

\begin{eqnarray}\label{MSSM_Yukawa_Renormalization}
&&\hspace*{-3mm} Y_{0U} = (Z_{H_u})^{-1/2} \left((Z_Q)^T\right)^{-1/2} Y_{U} (Z_U)^{-1/2};\qquad Y_{0D} = (Z_{H_d})^{-1/2} \left((Z_Q)^T\right)^{-1/2} Y_{D} (Z_D)^{-1/2};\nonumber\\
&&\hspace*{-3mm} Y_{0E} = (Z_{H_d})^{-1/2} \left((Z_L)^T\right)^{-1/2} Y_{E} (Z_E)^{-1/2};\qquad \bm{\mu}_0 = (Z_{H_u})^{-1/2} (Z_{H_d})^{-1/2} \bm{\mu},
\end{eqnarray}

\noindent
so that

\begin{eqnarray}\label{Yukawa_Derivatives}
&& \frac{d Y_U}{d\ln\mu} = \frac{1}{2} \Big(\gamma_{H_u} Y_U + (\gamma_{Q})^T Y_U + Y_U \gamma_U\Big);\qquad \frac{d Y_D}{d\ln\mu} = \frac{1}{2} \Big(\gamma_{H_d} Y_D + (\gamma_{Q})^T Y_D + Y_D \gamma_D\Big);\nonumber\\
&& \frac{d Y_E}{d\ln\mu} = \frac{1}{2} \Big(\gamma_{H_d} Y_E + (\gamma_{L})^T Y_E + Y_E \gamma_E\Big);\qquad \frac{d\bm{\mu}}{d\ln\mu} = \frac{1}{2}\Big(\gamma_{H_u} + \gamma_{H_d}\Big)\bm{\mu}.
\end{eqnarray}

It is important that the equations (\ref{MSSM_Yukawa_Renormalization}) and, therefore, the equations (\ref{Yukawa_Derivatives}) are valid in the HD+MSL scheme because in this scheme all renormalization constants contain only powers of $\ln\Lambda/\mu$, where $\Lambda$ is the dimensionful regularization parameter. Taking into account that the NSVZ relations are also satisfied for this renormalization prescription we see that all exact equations describing the renormalization of the MSSM parameters are satisfied in the HD+MSL scheme, the use of which is always assumed in what follows. Note that in the $\overline{\mbox{DR}}$ scheme the NSVZ equations (\ref{MSSM_Beta_NSVZ}) are not valid starting from the three-loop approximation for the $\beta$-functions (and the two-loop approximation for the anomalous dimensions), see \cite{Jack:2004ch} for details.

\section{All-loop RGIs for MSSM}
\hspace*{\parindent}\label{Section_RGI_MSSM}

The renormalization group equations (\ref{Yukawa_Derivatives}) for the Yukawa couplings can be multiplied by the corresponding inverse matrices. After that, it is possible to calculate traces of the resulting equations using the formula

\begin{equation}
\mbox{tr}\Big[M^{-1} \frac{dM}{d\ln\mu}\Big] = \frac{d}{d\ln\mu} \mbox{tr} \ln M = \frac{d}{d\ln\mu} \ln\mbox{det}\, M,
\end{equation}

\noindent
which can easily be proved if we present the matrix $M$ in the form $M=1+m$,

\begin{eqnarray}
&& \frac{d}{d\ln\mu}\mbox{tr}\ln M = \mbox{tr}\,\frac{d}{d\ln\mu}\ln(1+m) = \sum\limits_{n=1}^\infty \frac{(-1)^{n+1}}{n} \mbox{tr}\, \frac{d}{d\ln\mu} (m^n)\nonumber\\
&&\qquad\qquad = \sum\limits_{n=1}^\infty (-1)^{n+1} \mbox{tr}\Big(m^{n-1} \frac{dm}{d\ln\mu}\Big) = \mbox{tr}\Big(\frac{1}{1+m}\cdot \frac{dm}{\ln\mu}\Big) = \mbox{tr}\Big(M^{-1} \frac{dM}{d\ln\mu}\Big).\qquad
\end{eqnarray}

\noindent
Then (taking into account that the indices numerating generators range from 1 to 3) we see that the equations describing how the determinants of the Yukawa matrices depend on the renormalization point $\mu$ are written as

\begin{eqnarray}\label{Determinant_Equations}
&& \gamma_{\mbox{\scriptsize det}\,Y_U}\equiv \frac{d\ln\mbox{det}\, Y_U}{d\ln\mu}  = \mbox{tr}\Big[(Y_U)^{-1}\frac{d Y_U}{d\ln\mu}\Big] = \frac{1}{2} \Big(3\gamma_{H_u} + \mbox{tr}\big(\gamma_{Q} + \gamma_U\big)\Big);\quad\nonumber\\
&& \gamma_{\mbox{\scriptsize det}\,Y_D}\equiv \frac{d\ln\mbox{det}\, Y_D}{d\ln\mu} = \mbox{tr}\Big[(Y_D)^{-1}\frac{d Y_D}{d\ln\mu}\Big] = \frac{1}{2} \Big(3\gamma_{H_d} + \mbox{tr}\big(\gamma_{Q} + \gamma_D\big)\Big);\nonumber\\
&& \gamma_{\mbox{\scriptsize det}\,Y_E}\equiv \frac{d\ln\mbox{det}\, Y_E}{d\ln\mu} = \mbox{tr}\Big[(Y_E)^{-1}\frac{d Y_E}{d\ln\mu}\Big] = \frac{1}{2} \Big(3\gamma_{H_d} + \mbox{tr}\big(\gamma_{L} + \gamma_E\big)\Big).
\end{eqnarray}

\noindent
We will also need the last equation in (\ref{Yukawa_Derivatives}), which is convenient to present in the form

\begin{equation}\label{Mu_Renormalization}
\gamma_{\bm{\mu}} \equiv \frac{d\ln\bm{\mu}}{d\ln\mu} = \frac{1}{2}\Big(\gamma_{H_u} + \gamma_{H_d}\Big).
\end{equation}

It is also expedient to rewrite the equations (\ref{MSSM_Beta_NSVZ}) as

\begin{eqnarray}\label{MSSM_Beta_NSVZ_Equivalent}
&&\hspace*{-5mm} \frac{d}{d\ln\mu}\Big(\frac{1}{\alpha_1}\Big) = \frac{3}{5} \cdot \frac{1}{2\pi}\bigg[-11 + \mbox{tr}\Big(\frac{1}{6} \gamma_{Q} + \frac{4}{3} \gamma_{U} + \frac{1}{3} \gamma_{D} + \frac{1}{2} \gamma_{L}
+ \gamma_{E}\Big) + \frac{1}{2} \gamma_{H_u} + \frac{1}{2} \gamma_{H_d}\bigg];\quad\nonumber\\
&&\hspace*{-5mm} \frac{d}{d\ln\mu}\Big( \frac{1}{\alpha_2} + \frac{1}{\pi}\ln\alpha_2\Big) = \frac{1}{2\pi} \bigg[-1 + \mbox{tr}\Big(\frac{3}{2} \gamma_{Q} + \frac{1}{2} \gamma_{L}\Big) + \frac{1}{2} \gamma_{H_u} + \frac{1}{2} \gamma_{H_d}\bigg];\nonumber\\
&&\hspace*{-5mm} \frac{d}{d\ln\mu}\Big( \frac{1}{\alpha_3} + \frac{3}{2\pi}\ln \alpha_3\Big) = \frac{1}{2\pi} \bigg[3 + \mbox{tr} \Big(\gamma_{Q} + \frac{1}{2} \gamma_{U} + \frac{1}{2} \gamma_{D}\Big)\bigg].
\end{eqnarray}

The anomalous dimensions of the chiral matter superfields and $\bm{\mu}$ can be excluded from Eqs. (\ref{Determinant_Equations}), (\ref{Mu_Renormalization}), and (\ref{MSSM_Beta_NSVZ_Equivalent}), thereby obtaining a differential equation which contains only derivatives of the gauge and Yukawa couplings. Integrating it we obtain the expression $\mbox{RGI}_2$ in (\ref{First_Pair_Of_RGIs}), which vanishes after differentiating with respect to the renormalization point $\mu$. Alternatively, it is possible to construct the expression $\mbox{RGI}_1$ which does not explicitly depend on the scale $\mu$, but contains the parameter $\bm{\mu}$. The details of the calculation are presented in Appendix \ref{Appendix_Details}. Two resulting renormalization group invariants can be written in the form

\begin{eqnarray}\label{First_Pair_Of_RGIs}
&& \mbox{RGI}_1 = \frac{\bm{\mu}^{9/2}\,(\alpha_3)^3\,(\alpha_2)^{1/2}}{\big(\mbox{det}\, Y_E\big)^{1/2}\,\big(\mbox{det}\, Y_U\big)^{5/3}\, \big(\mbox{det}\, Y_D\big)^{7/6}}\, \exp\Big(\frac{2\pi}{\alpha_3} + \frac{\pi}{2\alpha_2} +\frac{5\pi}{6\alpha_1} \Big);\qquad\nonumber\\
&&\mbox{RGI}_2 = \frac{(\alpha_3)^3\, \mbox{det}\, Y_E\, \big(\mbox{det}\, Y_U \big)^{1/3}}{\mu^9\,\alpha_2\, \big(\mbox{det}\, Y_D \big)^{2/3}}\, \exp\Big(\frac{2\pi}{\alpha_3} -\frac{\pi}{\alpha_2} - \frac{5\pi}{3\alpha_1}\Big).
\end{eqnarray}

Instead of the renormalization group invariants $(\mbox{RGI}_1,\ \mbox{RGI}_2)$ it is possible to use the equivalent set $(\mbox{RGI}_3,\ \mbox{RGI}_4)$, where the expressions

\begin{eqnarray}\label{Second_Pair_Of_RGI}
&& \mbox{RGI}_3 \equiv \Big(\frac{\mbox{RGI}_1}{\mbox{RGI}_2}\Big)^{2/3} = \frac{\bm{\mu}^{3}\,\mu^6\,\alpha_2}{\big(\mbox{det}\, Y_E\big)\,\big(\mbox{det}\, Y_U\big)^{4/3}\, \big(\mbox{det}\, Y_D\big)^{1/3}}\, \exp\Big(\frac{\pi}{\alpha_2} +\frac{5\pi}{3\alpha_1} \Big);\qquad\nonumber\\
&& \mbox{RGI}_4 \equiv \big(\mbox{RGI}_1\big)^{2/3} \big(\mbox{RGI}_2\big)^{1/3} =
\frac{\bm{\mu}^{3}\,(\alpha_3)^3}{\mu^3\,\mbox{det}\, Y_U\, \mbox{det}\, Y_D}\, \exp\Big(\frac{2\pi}{\alpha_3} \Big)
\end{eqnarray}

\noindent
also have a rather simple form.

\section{RGIs for NMSSM}
\hspace*{\parindent}\label{Section_NMSSM}

The parameter $\bm{\mu}$ in the superpotential (\ref{MSSM_Superpotential}) should be of the order of the electroweak scale, which is impossible to explain in MSSM. The $\bm{\mu}$ problem can be solved in the Next-to-Minimal Supersymmetric Standard Model (NMSSM) \cite{Maniatis:2009re,Ellwanger:2009dp}, which contains an additional chiral matter superfield $S$. This superfield is a singlet with respect to $SU(3)\times SU(2)\times U(1)$. Then it is possible to replace the $\bm{\mu}$ term

\begin{equation}
\Delta W_{\mbox{\scriptsize MSSM}} = \bm{\mu} \left(H_{u1}\ H_{u2} \right)
\left(
\begin{array}{cc}
0 & 1\\
-1 & 0
\end{array}
\right)
\left(
\begin{array}{c}
H_{d1}\\ H_{d2}
\end{array}
\right)
\end{equation}

\noindent
by the gauge invariant expression

\begin{equation}
\Delta W_{\mbox{\scriptsize NMSSM}} = \lambda S \left(H_{u1}\ H_{u2} \right)
\left(
\begin{array}{cc}
0 & 1\\
-1 & 0
\end{array}
\right)
\left(
\begin{array}{c}
H_{d1}\\ H_{d2}
\end{array}
\right) + \frac{\kappa}{3} S^3,
\end{equation}

\noindent
in which $\lambda$ and $\kappa$ are new dimensionless couplings. In this case the effective value of $\bm{\mu}$ is equal to the vacuum expectation value of (the lowest component of) $S$ multiplied by $\lambda$ and can have an order of the electroweak scale. Due to the nonrenormalization of the superpotential, the anomalous dimensions of $\lambda$ and $\kappa$ satisfy the all-loop equations

\begin{equation}\label{NMSSM_Yuakawa}
\gamma_\kappa \equiv \frac{d\ln\kappa}{d\ln\mu} = \frac{3}{2}\gamma_S;\qquad \gamma_\lambda \equiv \frac{d\ln\lambda}{d\ln\mu} = \frac{1}{2}\Big(\gamma_S +\gamma_{H_u} + \gamma_{H_d}\Big),
\end{equation}

\noindent
where the derivatives with respect to $\ln\mu$ should be taken at fixed values of all bare couplings. From Eq. (\ref{NMSSM_Yuakawa}) we see that the sum which for MSSM gives the anomalous dimension $\gamma_{\bm{\mu}}$ can be expressed in terms of $\gamma_\lambda$ and $\gamma_\kappa$,

\begin{equation}
\frac{1}{2}\Big(\gamma_{H_u} + \gamma_{H_d}\Big) = \gamma_\lambda - \frac{1}{3}\gamma_\kappa.
\end{equation}

\noindent
Since $S$ is a singlet with respect to the gauge group, the NSVZ relations for NMSSM are the same as for MSSM (although the anomalous dimensions are different). Evidently, the equations describing the running of the Yukawa couplings $Y_E$, $Y_U$, and $Y_D$ also remain unchanged. That is why RGIs for NMSSM can be obtained from the ones for MSSM after the replacement

\begin{equation}\label{Replacement}
\bm{\mu}\ \to\ \lambda \kappa^{-1/3}.
\end{equation}

\noindent
The expression $\mbox{RGI}_2$ does not depend on $\bm{\mu}$ and, therefore, is also RGI for NMSSM. The other RGIs after the replacement (\ref{Replacement}) give

\begin{eqnarray}\label{NMSSM_RGI}
&& \mbox{RGI}_1\ \to\ \widetilde{\mbox{RGI}}_1 = \frac{\lambda^{9/2}\,(\alpha_3)^3\,(\alpha_2)^{1/2}}{\kappa^{3/2}\,\big(\mbox{det}\, Y_E\big)^{1/2}\,\big(\mbox{det}\, Y_U\big)^{5/3}\, \big(\mbox{det}\, Y_D\big)^{7/6}}\, \exp\Big(\frac{2\pi}{\alpha_3} + \frac{\pi}{2\alpha_2} +\frac{5\pi}{6\alpha_1} \Big);\qquad\nonumber\\
&& \mbox{RGI}_3\ \to\ \widetilde{\mbox{RGI}}_3 = \frac{\lambda^{3}\,\mu^6\,\alpha_2}{\kappa\, \big(\mbox{det}\, Y_E\big)\,\big(\mbox{det}\, Y_U\big)^{4/3}\, \big(\mbox{det}\, Y_D\big)^{1/3}}\, \exp\Big(\frac{\pi}{\alpha_2} +\frac{5\pi}{3\alpha_1} \Big);\qquad\nonumber\\
&& \mbox{RGI}_4\ \to\ \widetilde{\mbox{RGI}}_4 = \frac{\lambda^{3}\,(\alpha_3)^3}{\kappa\, \mu^3\,\mbox{det}\, Y_U\, \mbox{det}\, Y_D}\, \exp\Big(\frac{2\pi}{\alpha_3} \Big).
\end{eqnarray}

\noindent
Thus, for NMSSM there are also two independent invariants. As for MSSM, it is possible to use two convenient choices, $(\widetilde{\mbox{RGI}}_1,\,\mbox{RGI}_2)$ or $(\widetilde{\mbox{RGI}}_3,\,\widetilde{\mbox{RGI}}_4)$.

\section{The three-loop check}
\hspace*{\parindent}\label{Section_Verification}

To verify the renormalization group invariance of the expressions constructed in Sect. \ref{Section_RGI_MSSM}, we will use the three-loop results for MSSM $\beta$-functions and the two-loop results for the anomalous dimensions of the chiral matter superfields. In the HD+MSL and $\overline{\mbox{DR}}$ schemes they have been calculated in \cite{Haneychuk:2022qvu} and \cite{Jack:2004ch}, respectively. It is easier to check the invariance of the set $(\mbox{RGI}_3,\, \mbox{RGI}_4)$, which is what we will do. Differentiating $\ln(\mbox{RGI}_3)$ and $\ln(\mbox{RGI}_4)$ with respect to $\ln\mu$ we obtain the exact equations

\begin{eqnarray}\label{Third_Equation}
&& 0 = \Big(\frac{1}{\alpha_2} - \frac{\pi}{\alpha_2^2}\Big)\beta_2 - \frac{5\pi}{3\alpha_1^2} \beta_1 + 6 + 3\gamma_{\bm{\mu}} - \gamma_{\mbox{\scriptsize det}\,Y_E} - \frac{4}{3}\gamma_{\mbox{\scriptsize det}\,Y_U} - \frac{1}{3} \gamma_{\mbox{\scriptsize det}\,Y_D};\qquad\\
\label{Fourth_Equation}
&& 0 = \Big(\frac{3}{\alpha_3} - \frac{2\pi}{\alpha_3^2}\Big) \beta_3 -3 + 3\gamma_{\bm{\mu}} - \gamma_{\mbox{\scriptsize det}\,Y_U} - \gamma_{\mbox{\scriptsize det}\,Y_D}.
\end{eqnarray}

\noindent
The expressions entering these equations in the HD+MSL and $\overline{\mbox{DR}}$ schemes are presented in Appendices \ref{Appendix_RGFs_HD+MSL} and \ref{Appendix_RGFs_DR}, respectively. In the HD+MSL scheme they depend on the regularization parameters

\begin{eqnarray}\label{Regularization_Parameters}
&& A\equiv \int\limits_{0}^\infty dx\,\ln x\,\frac{d}{dx} \frac{1}{R(x)};\qquad a_{\varphi,3} \equiv \frac{M_{\varphi,3}}{\Lambda};\quad\ \ a_{\varphi,2} \equiv \frac{M_{\varphi,2}}{\Lambda};\qquad\nonumber\\
&& B\equiv \int\limits_{0}^\infty dx\,\ln x\,\frac{d}{dx} \frac{1}{F^2(x)};\quad\ \ a_3 \equiv \frac{M_3}{\Lambda};\qquad\quad a_2 \equiv \frac{M_2}{\Lambda};\qquad\quad a_1 \equiv \frac{M_1}{\Lambda}.\qquad
\end{eqnarray}

\noindent
Here the functions $R(x)$ and $F(x)$ are used for introducing higher covariant derivatives in the gauge and matter parts of the action, respectively.\footnote{The detailed description of the higher covariant derivative regularization for MSSM can be found in \cite{Korneev:2021zdz}.} They satisfy the conditions $R(0)=1$, $F(0)=1$ and should rapidly increase in the limit $x\to \infty$. After introducing higher derivatives in the classical action divergences remain only in the one-loop approximation. These remaining divergences are regularized by inserting the Pauli--Villars determinants into the generating functional \cite{Slavnov:1977zf}. According to \cite{Aleshin:2016yvj,Kazantsev:2017fdc}, in the case of a simple gauge group it is necessary to use two sets of the Pauli--Villars superfields. One of them has the mass $M_\varphi$ and cancels one-loop divergences coming from the quantum gauge superfield and ghosts, while the other has the mass $M$ and cancels the one-loop divergences coming from the matter superfields. For theories with multiple gauge couplings such sets should be introduced for each simple subgroup of the gauge group. For each $U(1)$ subgroup it is necessary to introduce only the latter the Pauli--Villars superfield. In particular, for MSMM we need the Pauli--Villars superfields with the mass $M_1$ for the subgroup $U(1)$. For the subgroups $SU(2)$ and $SU(3)$ the Pauli--Villars masses are $M_{\varphi,2}$, $M_{2}$ and $M_{\varphi,3}$, $M_3$, respectively. Due to the arbitrariness in the choice of the regulators $R$, $F$, and the parameters $a_{\varphi,K}$ and $a_K$ there is a class of the HD+MSL scheme, each of them being NSVZ.

Substituting the expressions for RGFs in the HD+MSL scheme (presented in Appendix \ref{Appendix_RGFs_HD+MSL}) into Eqs. (\ref{Third_Equation}) and (\ref{Fourth_Equation}) we see that in the considered approximation these equations are valid independently of the values of the regularization parameters (\ref{Regularization_Parameters}),

\begin{eqnarray}
&&\hspace*{-5mm} \bigg[\Big(\frac{1}{\alpha_2} - \frac{\pi}{\alpha_2^2}\Big)\beta_2 - \frac{5\pi}{3\alpha_1^2} \beta_1 + 6 + 3\gamma_{\bm{\mu}} - \gamma_{\mbox{\scriptsize det}\,Y_E} - \frac{4}{3}\gamma_{\mbox{\scriptsize det}\,Y_U} - \frac{1}{3} \gamma_{\mbox{\scriptsize det}\,Y_D}\bigg]_{\mbox{\scriptsize HD+MSL}}\nonumber\\
&&\hspace*{-5mm}\qquad\qquad\qquad\qquad\qquad\qquad\qquad\qquad\qquad\qquad\qquad\quad\ \, = O\Big(\alpha^3,\alpha^2 Y^2,\alpha Y^4, Y^6\Big);\vphantom{\frac{1}{2}}\\
&& \vphantom{1}\nonumber\\
&&\hspace*{-5mm} \bigg[\Big(\frac{3}{\alpha_3} - \frac{2\pi}{\alpha_3^2}\Big) \beta_3 -3 + 3\gamma_{\bm{\mu}} - \gamma_{\mbox{\scriptsize det}\,Y_U} - \gamma_{\mbox{\scriptsize det}\,Y_D}\bigg]_{\mbox{\scriptsize HD+MSL}} =  O\Big(\alpha^3,\alpha^2 Y^2,\alpha Y^4, Y^6\Big).\quad
\end{eqnarray}

\noindent
Therefore, the expressions (\ref{Second_Pair_Of_RGI}) (and, therefore, the expressions (\ref{First_Pair_Of_RGIs})) in the HD+MSL scheme(s) are also independent of $\mu$ in the considered order.

However, substituting RGFs obtained in the $\overline{\mbox{DR}}$ scheme we see that Eqs. (\ref{Third_Equation}) and (\ref{Fourth_Equation}) are not satisfied,

\begin{eqnarray}
&&\hspace*{-5mm} \bigg[\Big(\frac{1}{\alpha_2} - \frac{\pi}{\alpha_2^2}\Big)\beta_2 - \frac{5\pi}{3\alpha_1^2} \beta_1 + 6 + 3\gamma_{\bm{\mu}} - \gamma_{\mbox{\scriptsize det}\,Y_E} - \frac{4}{3}\gamma_{\mbox{\scriptsize det}\,Y_U} - \frac{1}{3} \gamma_{\mbox{\scriptsize det}\,Y_D}\bigg]_{\overline{\mbox{\scriptsize DR}}}\nonumber\\
&&\hspace*{-5mm} = \frac{1}{2\pi^2}\Big(\frac{1243\alpha_1^2}{400}+\frac{17\alpha_2^2}{16} - 5\alpha_3^2\Big)
+\frac{1}{16\pi^3} \mbox{tr}(Y_U Y_U^+) \Big(\frac{143\alpha_1}{180}+\frac{11\alpha_2}{4}+\frac{44\alpha_3}{9}\Big)
\nonumber\\
&&\hspace*{-5mm} +\frac{1}{16\pi^3} \mbox{tr}(Y_D Y_D^+) \Big(\frac{14\alpha_1}{45}+2\alpha_2+\frac{32\alpha_3}{9}\Big)
+\frac{1}{16\pi^3} \mbox{tr}(Y_E Y_E^+) \Big(\frac{9\alpha_1}{10}+\frac{3\alpha_2}{2}\Big)
\nonumber\\
&&\hspace*{-5mm} -\frac{1}{(16\pi^2)^2}\bigg[\, 11\,\mbox{tr}\Big((Y_{U} Y_{U}^+)^2\Big) + 8\,\mbox{tr}\Big((Y_{D} Y_{D}^+)^2\Big) + 6\,\mbox{tr}\Big((Y_{E} Y_{E}^+)^2\Big)+ \frac{19}{3} \mbox{tr}\Big(Y_{D} Y_{D}^+ Y_{U} Y_{U}^+\Big)
\nonumber\\
&&\hspace*{-5mm}  + 11\Big(\mbox{tr}(Y_{U} Y_{U}^+)\Big)^2 + 8\Big(\mbox{tr}(Y_{D} Y_{D}^+)\Big)^2 + 2\Big(\mbox{tr}(Y_{E} Y_{E}^+)\Big)^2 + \frac{26}{3} \mbox{tr}\Big(Y_{E} Y_{E}^+\Big)\,\mbox{tr}\Big(Y_{D} Y_{D}^+\Big) \bigg]\nonumber\\
&&\hspace*{-5mm} + O\Big(\alpha^3,\alpha^2 Y^2,\alpha Y^4, Y^6\Big) \ne O\Big(\alpha^3,\alpha^2 Y^2,\alpha Y^4, Y^6\Big);\vphantom{\frac{1}{2}}\\
&& \vphantom{1}\nonumber\\
&&\hspace*{-5mm} \bigg[\Big(\frac{3}{\alpha_3} - \frac{2\pi}{\alpha_3^2}\Big) \beta_3 -3 + 3\gamma_{\bm{\mu}} - \gamma_{\mbox{\scriptsize det}\,Y_U} - \gamma_{\mbox{\scriptsize det}\,Y_D}\bigg]_{\overline{\mbox{\scriptsize DR}}}
= \frac{1}{2\pi^2}\Big(\frac{363\alpha_1^2}{400}+\frac{9\alpha_2^2}{16} - \frac{21\alpha_3^2}{8}\Big) \nonumber\\
&&\hspace*{-5mm} +\frac{1}{16\pi^3} \mbox{tr}(Y_U Y_U^+) \Big(\frac{13\alpha_1}{30}+\frac{3\alpha_2}{2}+\frac{8\alpha_3}{3}\Big) +\frac{1}{16\pi^3} \mbox{tr}(Y_D Y_D^+) \Big(\frac{7\alpha_1}{30}+\frac{3\alpha_2}{2}+\frac{8\alpha_3}{3}\Big)
\nonumber\\
&&\hspace*{-5mm} -\frac{1}{(16\pi^2)^2}\bigg[\, 6\,\mbox{tr}\Big((Y_{U} Y_{U}^+)^2\Big) + 6\,\mbox{tr}\Big((Y_{D} Y_{D}^+)^2\Big) + 6\Big(\mbox{tr}(Y_{U} Y_{U}^+)\Big)^2 + 6\Big(\mbox{tr}(Y_{D} Y_{D}^+)\Big)^2 \nonumber\\
&&\hspace*{-5mm} + 2\,\mbox{tr}\Big(Y_{E} Y_{E}^+\Big)\,\mbox{tr}\Big(Y_{D} Y_{D}^+\Big) + 4\,\mbox{tr}\Big(Y_{D} Y_{D}^+ Y_{U} Y_{U}^+\Big)\bigg]  + O\Big(\alpha^3,\alpha^2 Y^2,\alpha Y^4, Y^6\Big)\nonumber\\
&&\hspace*{-5mm} \ne O\Big(\alpha^3,\alpha^2 Y^2,\alpha Y^4, Y^6\Big).
\end{eqnarray}

\noindent
Certainly, this occurs in such an approximation where the scheme dependence becomes essential.

\section{Conclusion}
\hspace*{\parindent}

In this paper we have demonstrated that for MSSM it is possible to construct two independent scale invariant expressions from the gauge couplings, Yukawa couplings, $\bm{\mu}$ parameter, and the renormalization point $\mu$. One of them can be chosen independent of the parameter $\bm{\mu}$. The derivation is based on the all-loop exact equations following from the NSVZ $\beta$-functions and the nonrenormalization of the superpotential. That is why the renormalization group invariance takes place only for certain renormalization prescriptions in which these equations are satisfied. In particular, it should be valid in all loops in the HD+MSL scheme(s), when a theory is regularized by higher covariant derivatives and the renormalization is made with the help of minimal subtractions of logarithms. This statement has been verified in the three-loop approximation for the $\beta$-functions and two-loop approximaion for the anomalous dimensions (of the Yukawa couplings and the $\bm{\mu}$ parameter). In the $\overline{\mbox{DR}}$ scheme the expressions constructed in this paper are not RGIs starting from such orders of the perturbation theory where the scheme dependence of RGFs becomes essential. The results can also be extended to the case of NMSSM.

\appendix

\section*{Appendix}

\section{Detailed derivation of RGIs for MSSM}
\hspace{\parindent}\label{Appendix_Details}

We start with the system of all-order exact equations (\ref{Determinant_Equations}), (\ref{Mu_Renormalization}), and (\ref{MSSM_Beta_NSVZ_Equivalent}) and exclude from them the expressions $\gamma_{H_u}$ and $\gamma_{H_d}$. As the result, we obtain the system of differential equations

\begin{eqnarray}
&& \frac{d}{d\ln\mu}\Big( \frac{2\pi}{\alpha_3} + 3\ln \alpha_3\Big) = 3 + \mbox{tr} \Big(\gamma_{Q} + \frac{1}{2} \gamma_{U} + \frac{1}{2} \gamma_{D}\Big);\nonumber\\
&& \frac{d}{d\ln\mu}\Big( \frac{2\pi}{\alpha_2} + 2 \ln\alpha_2 - \ln\bm{\mu}\Big) = -1 + \mbox{tr}\Big(\frac{3}{2} \gamma_{Q} + \frac{1}{2} \gamma_{L}\Big);\nonumber\\
&& \frac{d}{d\ln\mu}\Big(\frac{5}{3}\cdot\frac{2\pi}{\alpha_1} - \ln\bm{\mu}\Big) = -11 + \mbox{tr}\Big(\frac{1}{6} \gamma_{Q} + \frac{4}{3} \gamma_{U} + \frac{1}{3} \gamma_{D} + \frac{1}{2} \gamma_{L}
+ \gamma_{E}\Big); \quad\nonumber\\
&& \frac{d}{d\ln\mu}\Big(\ln\mbox{det}\, Y_D + \ln\mbox{det}\, Y_U - 3\ln\bm{\mu}\Big) = \frac{1}{2} \mbox{tr}\Big(2\gamma_{Q} +\gamma_U +\gamma_D\Big);\nonumber\\
&& \frac{d}{d\ln\mu}\Big(\ln\mbox{det}\, Y_E - \ln\mbox{det}\, Y_D\Big) = \frac{1}{2} \mbox{tr}\Big(\gamma_L +\gamma_E -\gamma_Q -\gamma_D\Big).
\end{eqnarray}

\noindent
After that, we exclude the expressions $\mbox{tr}(\gamma_{L})$ and $\mbox{tr}(\gamma_{E})$. The resulting equations contain the anomalous dimensions $\gamma_Q$, $\gamma_D$, and $\gamma_U$ only in the combination $\mbox{tr}(2\gamma_{Q}+\gamma_{U}+\gamma_D)$,

\begin{eqnarray}
&& \frac{d}{d\ln\mu}\Big( \frac{2\pi}{\alpha_3} + 3\ln \alpha_3\Big) = 3 + \frac{1}{2}\mbox{tr} \Big(2\gamma_{Q} + \gamma_{U} + \gamma_{D}\Big);\nonumber\\
&& \frac{d}{d\ln\mu}\Big(\frac{2\pi}{\alpha_2} + 2 \ln\alpha_2 +\frac{5}{3}\cdot\frac{2\pi}{\alpha_1} - 2\ln\bm{\mu} -2\ln\mbox{det}\, Y_E +2\ln\mbox{det}\, Y_D\Big)\nonumber\\
&&\qquad\qquad\qquad\qquad\qquad\qquad\qquad\qquad\qquad\qquad = -12 + \frac{4}{3}\, \mbox{tr}\Big(2\gamma_{Q} + \gamma_{U} +\gamma_{D} \Big); \qquad\nonumber\\
&& \frac{d}{d\ln\mu}\Big(\ln\mbox{det}\, Y_D + \ln\mbox{det}\, Y_U - 3\ln\bm{\mu}\Big) = \frac{1}{2}\, \mbox{tr}\Big(2\gamma_{Q} +\gamma_U +\gamma_D\Big).
\end{eqnarray}

\noindent
This implies that after excluding all anomalous dimensions of the MSSM chiral matter superfields, we obtain the system of two differential equations,

\begin{eqnarray}
&& \frac{d}{d\ln\mu}\Big(\frac{8}{3}\cdot \frac{2\pi}{\alpha_3} +8\ln \alpha_3 - \frac{2\pi}{\alpha_2} - 2 \ln\alpha_2 -\frac{5}{3}\cdot\frac{2\pi}{\alpha_1}\nonumber\\
&&\qquad\qquad\qquad\qquad\qquad\qquad + 2\ln\bm{\mu} +2\ln\mbox{det}\, Y_E -2\ln\mbox{det}\, Y_D - 20\ln\mu\Big) = 0;\qquad \nonumber\\
&& \frac{d}{d\ln\mu}\Big( \frac{2\pi}{\alpha_3} + 3\ln \alpha_3 - \ln\mbox{det}\, Y_D - \ln\mbox{det}\, Y_U + 3\ln\bm{\mu} -3\ln\mu\Big) = 0.
\end{eqnarray}

Next, there are two options. Namely, it is possible either to construct an equation which does not explicitly depend on the scale $\mu$, or does not contain the MSSM parameter $\bm{\mu}$. In the former case we
obtain the differential equation

\begin{eqnarray}
&& \frac{d}{d\ln\mu}\Big(\frac{2\pi}{\alpha_3} + 3\ln \alpha_3 + \frac{\pi}{2\alpha_2} + \frac{1}{2} \ln\alpha_2 + \frac{5\pi}{6\alpha_1}\nonumber\\
&&\qquad\qquad\qquad\qquad -\frac{1}{2}\ln\mbox{det}\, Y_E - \frac{7}{6}\ln\mbox{det}\, Y_D - \frac{5}{3}\ln\mbox{det}\, Y_U + \frac{9}{2}\ln\bm{\mu} \Big) = 0, \qquad
\end{eqnarray}

\noindent
from which we conclude that the expression

\begin{equation}
\frac{\bm{\mu}^{9/2}\,(\alpha_3)^3\,(\alpha_2)^{1/2}}{\big(\mbox{det}\, Y_E\big)^{1/2}\,\big(\mbox{det}\, Y_U\big)^{5/3}\, \big(\mbox{det}\, Y_D\big)^{7/6}}\, \exp\Big(\frac{2\pi}{\alpha_3} + \frac{\pi}{2\alpha_2} +\frac{5\pi}{6\alpha_1} \Big) \equiv \mbox{RGI}_1
\end{equation}

\noindent
is independent of $\mu$.

Alternatively, it is possible to exclude the parameter $\bm{\mu}$ present in the MSSM superpotential. The resulting differential equation can be written in the form

\begin{eqnarray}
&& \frac{d}{d\ln\mu}\Big(\frac{2\pi}{\alpha_3} +3\ln \alpha_3 -\frac{\pi}{\alpha_2} - \ln\alpha_2 -\frac{5\pi}{3\alpha_1}\nonumber\\
&&\qquad\qquad\qquad\qquad +\ln\mbox{det}\, Y_E -\frac{2}{3} \ln\mbox{det}\, Y_D +\frac{1}{3}\ln\mbox{det}\, Y_U -9\ln\mu\Big) = 0.\qquad
\end{eqnarray}

\noindent
Integrating it we obtain the second RGI

\begin{equation}
\frac{(\alpha_3)^3\, \mbox{det}\, Y_E\, \big(\mbox{det}\, Y_U \big)^{1/3}}{\mu^9\,\alpha_2\, \big(\mbox{det}\, Y_D \big)^{2/3}}\, \exp\Big(\frac{2\pi}{\alpha_3} -\frac{\pi}{\alpha_2} - \frac{5\pi}{3\alpha_1}\Big)
\equiv \mbox{RGI}_2,
\end{equation}

\noindent
which, however, explicitly depends on the renormalization point $\mu$.

\section{Expressions for RGFs in the HD+MSL scheme}
\hspace*{\parindent}\label{Appendix_RGFs_HD+MSL}

The expressions for all MSSM $\beta$-functions in the three-loop approximation and all anomalous dimensions of the chiral matter superfields in the two-loop approximation for an arbitrary renormalization prescription supplementing the higher covariant derivative regularization have been calculated in \cite{Haneychuk:2022qvu}. In particular, it is easy to obtain the results in the HD+MSL scheme setting all finite constants to 0 (or, equivalently, replacing $\alpha_{0i}\to \alpha_i$, $Y_{0i}\to Y_i$ in RGFs defined in terms of the bare couplings). After that, the expressions for the three-loop MSSM $\beta$-functions in the HD+MSL scheme take the form

\begin{eqnarray}\label{Beta1_HD+MSL}
&&\hspace*{-5mm} \frac{\beta_1(\alpha,Y)}{\alpha_{1}^2} = - \frac{1}{2\pi}\cdot \frac{3}{5} \bigg\{ -11 -\frac{199\alpha_{1}}{60\pi} -\frac{9\alpha_{2}}{4\pi} -\frac{22\alpha_{3}}{3\pi}
+ \frac{1}{8\pi^2} \mbox{tr}\Big(\frac{13}{3} Y_{U} Y_{U}^+ + \frac{7}{3} Y_{D} Y_{D}^+ + 3 Y_{E} Y_{E}^+ \Big) \nonumber\\
&&\hspace*{-5mm}
+ \frac{1}{2\pi^2}\bigg[\, \frac{5131\alpha_{1}^2}{3600} + \frac{27\alpha_{2}^2}{16} + \frac{88\alpha_{3}^2}{9} + \frac{23\alpha_{1}\alpha_{2}}{40} + \frac{137\alpha_{1}\alpha_{3}}{45} + \alpha_{2}\alpha_{3}
+ \frac{2189\alpha_{1}^2}{100}\Big(\ln a_{1}+1+\frac{A}{2}\Big)
\nonumber\\
&&\hspace*{-5mm}
+ \frac{9\alpha_{2}^2}{4}\Big(7\ln a_{2}-6\ln a_{\varphi,2}+1+\frac{A}{2}\Big) -22 \alpha_{3}^2 \Big(3\ln a_{\varphi,3}-2\ln a_{3}+1+\frac{A}{2}\Big) \bigg]
+ \frac{1}{8\pi^3}\mbox{tr}\Big(Y_{U} Y_{U}^+\Big) \nonumber\\
&&\hspace*{-5mm}
\times
\bigg[\,2\alpha_{2} + 2\alpha_{3} + (B-A)\Big(\frac{169\alpha_{1}}{180} + \frac{13\alpha_{2}}{4} + \frac{52\alpha_{3}}{9}\Big)\bigg]
+ \frac{1}{8\pi^3}\mbox{tr}\Big(Y_{D} Y_{D}^+\Big)\bigg[\,\frac{\alpha_{2}}{2} + 2\alpha_{3} + (B-A) \nonumber\\
&&\hspace*{-5mm} \times\Big(\frac{49\alpha_{1}}{180}
+ \frac{7\alpha_{2}}{4} + \frac{28\alpha_{3}}{9}\Big)\bigg]
+ \frac{1}{8\pi^3}\mbox{tr}\Big(Y_{E} Y_{E}^+\Big)\bigg[\,\frac{3\alpha_{2}}{2} + (B-A)\Big(\frac{27\alpha_{1}}{20} + \frac{9\alpha_{2}}{4} \Big)\bigg]
- \frac{1}{(8\pi^2)^2}\bigg[\,\frac{15}{4} \nonumber\\
&&\hspace*{-5mm}
\times\, \mbox{tr}\Big((Y_{U} Y_{U}^+)^2\Big) + \frac{11}{4} \mbox{tr}\Big((Y_{D} Y_{D}^+)^2\Big)
+ \frac{9}{4} \mbox{tr}\Big((Y_{E} Y_{E}^+)^2\Big) + \frac{19}{6}\mbox{tr}\Big(Y_{D} Y_{D}^+ Y_{U} Y_{U}^+\Big) + \frac{17}{4} \Big(\mbox{tr}(Y_{U} Y_{U}^+)\Big)^2 \nonumber\\
&&\hspace*{-5mm}
+ \frac{5}{4} \Big(\mbox{tr}(Y_{D} Y_{D}^+)\Big)^2  + \frac{5}{4} \Big(\mbox{tr}(Y_{E} Y_{E}^+)\Big)^2
+ \frac{25}{6} \mbox{tr}\Big(Y_{E} Y_{E}^+\Big)\,\mbox{tr}\Big(Y_{D} Y_{D}^+\Big)  \bigg]\bigg\} + O(\alpha^3,\alpha^2 Y^2, \alpha Y^4, Y^6);\vphantom{\frac{1}{2}}\nonumber\\
&&\vphantom{1}\\
\label{Beta2_HD+MSL}
&&\hspace*{-5mm} \frac{\beta_2(\alpha,Y)}{\alpha_{2}^2} = - \frac{1}{2\pi} \bigg\{-1  -\frac{9\alpha_{1}}{20\pi}- \frac{25\alpha_{2}}{4\pi} -\frac{6\alpha_{3}}{\pi} + \frac{1}{8\pi^2}\,
\mbox{tr}\Big(3\, Y_{U} Y_{U}^+ + 3\, Y_{D} Y_{D}^+ + Y_{E} Y_{E}^+ \Big) \nonumber\\
&&\hspace*{-5mm} + \frac{1}{2\pi^2}\bigg[\frac{23\alpha_{1}^2}{400} - \frac{137\alpha_{2}^2}{16} + 8 \alpha_{3}^2 - \frac{9\alpha_{1}\alpha_{2}}{40} + \frac{\alpha_{1}\alpha_{3}}{5} - 3\alpha_{2}\alpha_{3}
+ \frac{297\alpha_{1}^2}{100}\Big(\ln a_{1} + 1 + \frac{A}{2} \Big) + \frac{21\alpha_{2}^2}{4}\nonumber\\
&&\hspace*{-5mm} \times \Big(7\ln a_{2} -6\ln a_{\varphi,2} + 1 + \frac{A}{2} \Big) -18\alpha_{3}^2 \Big(3\ln a_{\varphi,3} - 2\ln a_{3} + 1 + \frac{A}{2}\Big)\bigg]
+ \frac{1}{8\pi^3}\mbox{tr}\Big(Y_{U} Y_{U}^+\Big) \nonumber\\
&&\hspace*{-5mm}
\times\bigg[\,\frac{2\alpha_{1}}{5} + 3\alpha_{2} + 2\alpha_{3} + (B-A)\Big(\frac{13\alpha_{1}}{20} + \frac{9\alpha_{2}}{4} + 4\alpha_{3}\Big)\bigg]
+ \frac{1}{8\pi^3}\mbox{tr}\Big(Y_{D} Y_{D}^+\Big)\bigg[\,\frac{\alpha_{1}}{10}+3\alpha_{2} +2\alpha_{3} \nonumber\\
&&\hspace*{-5mm} + (B-A) \Big(\frac{7\alpha_{1}}{20} + \frac{9\alpha_{2}}{4} + 4\alpha_{3}\Big)\bigg] + \frac{1}{8\pi^3}\mbox{tr}\Big(Y_{E} Y_{E}^+\Big)\bigg[\,\frac{3\alpha_{1}}{10} + \alpha_{2} + (B-A)\Big(\frac{9\alpha_{1}}{20} + \frac{3\alpha_{2}}{4}\Big)\bigg]\nonumber\\
&&\hspace*{-5mm}
- \frac{1}{(8\pi^2)^2}\bigg[\, \frac{15}{4}\,\mbox{tr}\Big((Y_{U} Y_{U}^+)^2\Big) + \frac{15}{4}\,\mbox{tr}\Big((Y_{D} Y_{D}^+)^2\Big)  + \frac{5}{4}\,\mbox{tr}\Big((Y_{E} Y_{E}^+)^2\Big)
+ \frac{9}{4}\Big(\mbox{tr}(Y_{U} Y_{U}^+)\Big)^2 \nonumber\\
&&\hspace*{-5mm} + \frac{9}{4}\Big(\mbox{tr}(Y_{D} Y_{D}^+)\Big)^2
+ \frac{1}{4}\Big(\mbox{tr}(Y_{E} Y_{E}^+)\Big)^2
+ \frac{3}{2} \mbox{tr}\Big(Y_{E} Y_{E}^+\Big)
\,\mbox{tr}\Big(Y_{D} Y_{D}^+\Big)
+ \frac{3}{2}\mbox{tr}\Big(Y_{D} Y_{D}^+ Y_{U} Y_{U}^+\Big) \bigg] \bigg\}\nonumber\\
&&\hspace*{-5mm} + O(\alpha^3,\alpha^2 Y^2, \alpha Y^4, Y^6);\\
&&\vphantom{1}\nonumber\\
\label{Beta3_HD+MSL}
&&\hspace*{-5mm} \frac{\beta_3(\alpha,Y)}{\alpha_{3}^2} = - \frac{1}{2\pi} \bigg\{3 -\frac{11\alpha_{1}}{20\pi} -\frac{9\alpha_{2}}{4\pi} -\frac{7\alpha_{3}}{2\pi}
+ \frac{1}{8\pi^2}\, \mbox{tr}\Big(2\, Y_{U} Y_{U}^+ + 2\, Y_{D} Y_{D}^+\Big) + \frac{1}{2\pi^2}\bigg[ \frac{137\alpha_{1}^2}{1200}\nonumber\\
&&\hspace*{-5mm}  + \frac{27\alpha_{2}^2}{16} + \frac{\alpha_{3}^2}{6} + \frac{3\alpha_{1}\alpha_{2}}{40} - \frac{11\alpha_{1}\alpha_{3}}{60}
-\frac{3\alpha_{2}\alpha_{3}}{4} + \frac{363\alpha_{1}^2}{100}\Big(\ln a_{1} + 1 +\frac{A}{2} \Big) + \frac{9\alpha_{2}^2}{4} \Big(-6\ln a_{\varphi,2}\nonumber\\
&&\hspace*{-5mm}  + 7\ln a_{2} + 1 +\frac{A}{2}\Big) - 24\alpha_{3}^2 \Big(3\ln a_{\varphi,3} -2\ln a_3 + 1 +\frac{A}{2} \Big) \bigg] + \frac{1}{8\pi^3}\mbox{tr}\Big(Y_{U} Y_{U}^+\Big)\bigg[\frac{3\alpha_{1}}{20} + \frac{3\alpha_{2}}{4}\nonumber\\
&&\hspace*{-5mm} + 3\alpha_{3} + (B-A)\Big(\frac{13\alpha_{1}}{30} + \frac{3\alpha_{2}}{2} + \frac{8\alpha_{3}}{3}\Big)\bigg] + \frac{1}{8\pi^3}\mbox{tr}\Big(Y_{D} Y_{D}^+\Big)\bigg[\frac{3\alpha_{1}}{20} + \frac{3\alpha_{2}}{4} + 3\alpha_{3} + (B-A)
\nonumber\\
&&\hspace*{-5mm} \times \Big(\frac{7\alpha_{1}}{30} + \frac{3\alpha_{2}}{2} + \frac{8\alpha_{3}}{3}\Big)\bigg] - \frac{1}{(8\pi^2)^2}\bigg[\, \frac{3}{2}\mbox{tr}\Big((Y_{U} Y_{U}^+)^2\Big) + \frac{3}{2}\mbox{tr}\Big((Y_{D} Y_{D}^+)^2\Big) + 3\Big(\mbox{tr}(Y_{U} Y_{U}^+)\Big)^2 \nonumber\\
&&\hspace*{-5mm} + 3\Big(\mbox{tr}(Y_{D} Y_{D}^+)\Big)^2 + \mbox{tr}\Big(Y_{E} Y_{E}^+\Big)\,\mbox{tr}\Big(Y_{D} Y_{D}^+\Big) + \mbox{tr}\Big(Y_{D} Y_{D}^+ Y_{U} Y_{U}^+\Big)\bigg]\bigg\}
+ O(\alpha^3,\alpha^2 Y^2, \alpha Y^4, Y^6).\qquad\nonumber\\
\end{eqnarray}

Similarly, starting from the expressions for the anomalous dimensions of the chiral matter superfields presented in \cite{Haneychuk:2022qvu}, it is possible to construct expressions for the anomalous dimensions entering Eqs. (\ref{Third_Equation}) and (\ref{Fourth_Equation}) in the two-loop approximation,

\begin{eqnarray}\label{Gamma_DetYU_HD+MSL}
&&\hspace*{-5mm} \gamma_{\mbox{\scriptsize det}\,Y_U}(\alpha,Y) = \frac{1}{2}\Big(3\gamma_{H_u}(\alpha,Y)  + \mbox{tr}\,\gamma_{Q}(\alpha,Y) + \mbox{tr}\,\gamma_U(\alpha,Y) \Big)\nonumber\\
&&\hspace*{-5mm} = -\frac{13\alpha_1}{20\pi} - \frac{9\alpha_2}{4\pi} - \frac{4\alpha_3}{\pi} + \frac{1}{16\pi^2}\, \mbox{tr}\Big(12\, Y_U Y_U^+ + Y_D Y_D^+\Big) + \frac{1}{2\pi^2}\bigg[\frac{169\alpha_1^2}{1200} + \frac{27\alpha_2^2}{16} + \frac{16\alpha_3^2}{3}\nonumber\\
&&\hspace*{-5mm} + \frac{3\alpha_1\alpha_2}{8} + \frac{17\alpha_1\alpha_3}{15} + 3\alpha_2\alpha_3 -\frac{27\alpha_2^2}{2}\Big(\ln a_{\varphi,2}+1+\frac{A}{2}\Big) - 36\alpha_3^2 \Big(\ln a_{\varphi,3}+1+\frac{A}{2}\Big)
\nonumber\\
&&\hspace*{-5mm}
+ \frac{429\alpha_1^2}{100}\Big(\ln a_{1}+1+\frac{A}{2}\Big) + \frac{63\alpha_2^2}{4}\Big(\ln a_{2}+1+\frac{A}{2}\Big) + 24\alpha_3^2 \Big(\ln a_{3}+1+\frac{A}{2}\Big) \bigg] + \frac{1}{16\pi^3}\nonumber\\
&&\hspace*{-5mm} \times\, \mbox{tr}(Y_{U} Y_{U}^+) \bigg[\frac{7\alpha_1}{10}+\frac{3\alpha_2}{2}+12\alpha_3 + (B-A)\Big(\frac{13\alpha_1}{5} + 9\alpha_2 + 16\alpha_3\Big)\bigg]
+ \frac{1}{16\pi^3} \mbox{tr}(Y_{D} Y_{D}^+)\nonumber\\
&&\hspace*{-5mm} \times\,\bigg[\frac{\alpha_{1}}{10} + (B-A)\Big(\frac{7\alpha_{1}}{60} +\frac{3\alpha_{2}}{4}+ \frac{4\alpha_{3}}{3} \Big) \bigg] - \frac{1}{(16\pi^2)^2}\bigg[\, 31\,\mbox{tr}\Big((Y_{U}Y_{U}^+)^2\Big) + 2\,\mbox{tr}\Big((Y_{D}Y_{D}^+)^2\Big)\nonumber\\
&&\hspace*{-5mm} +11\,\mbox{tr}\Big(Y_{D} Y_{D}^+ Y_{U} Y_{U}^+\Big) + 9 \Big(\mbox{tr}(Y_{U} Y_{U}^+)\Big)^2 + 3\Big(\mbox{tr}(Y_{D} Y_{D}^+)\Big)^2
+ \mbox{tr}(Y_{E} Y_{E}^+)\,\mbox{tr}(Y_{D} Y_{D}^+)
\bigg]\nonumber\\
&&\hspace*{-5mm} + O\Big(\alpha^3,\alpha^2 Y^2,\alpha Y^4, Y^6\Big);\\
&&\vphantom{1}\nonumber\\
\label{Gamma_DetYD_HD+MSL}
&&\hspace*{-5mm} \gamma_{\mbox{\scriptsize det}\,Y_D}(\alpha,Y) = \frac{1}{2}\Big(3\gamma_{H_d}(\alpha,Y)  + \mbox{tr}\,\gamma_{Q}(\alpha,Y) + \mbox{tr}\,\gamma_D(\alpha,Y) \Big)\nonumber\\
&&\hspace*{-5mm} = -\frac{7\alpha_1}{20\pi} - \frac{9\alpha_2}{4\pi} - \frac{4\alpha_3}{\pi}
+ \frac{1}{16\pi^2}\, \mbox{tr}\Big( Y_U Y_U^+ + 12\,Y_D Y_D^+ + 3\,Y_E Y_E^+\Big) + \frac{1}{2\pi^2}\bigg[\frac{49\alpha_1^2}{1200}
+ \frac{27\alpha_2^2}{16} \nonumber\\
&&\hspace*{-5mm} + \frac{16\alpha_3^2}{3} + \frac{3\alpha_1\alpha_2}{8} + \frac{\alpha_1\alpha_3}{3} + 3\alpha_2\alpha_3
-\frac{27\alpha_2^2}{2}\Big(\ln a_{\varphi,2}+1+\frac{A}{2}\Big) - 36\alpha_3^2 \Big(\ln a_{\varphi,3}+1+\frac{A}{2}\Big)
\nonumber\\
&&\hspace*{-5mm}
+ \frac{231\alpha_1^2}{100}\Big(\ln a_{1}+1+\frac{A}{2}\Big) + \frac{63\alpha_2^2}{4}\Big(\ln a_{2}+1+\frac{A}{2}\Big) + 24\alpha_3^2 \Big(\ln a_{3}+1+\frac{A}{2}\Big) \bigg] + \frac{1}{16\pi^3}\nonumber\\
&&\hspace*{-5mm} \times\, \mbox{tr}(Y_{U} Y_{U}^+) \bigg[\frac{\alpha_{1}}{5} + (B-A)\Big(\frac{13\alpha_{1}}{60} +\frac{3\alpha_{2}}{4} + \frac{4\alpha_{3}}{3} \Big)\bigg]
+ \frac{1}{16\pi^3} \mbox{tr}(Y_{D} Y_{D}^+) \bigg[-\frac{\alpha_{1}}{10} + \frac{3\alpha_2}{2} \nonumber\\
&&\hspace*{-5mm} + 12\alpha_3 + (B-A)\Big(\frac{7\alpha_{1}}{5} + 9\alpha_{2} + 16\alpha_{3} \Big) \bigg]
+ \frac{1}{16\pi^3} \mbox{tr}\Big(Y_{E} Y_{E}^+\Big)\bigg[\frac{9\alpha_{1}}{10} + (B-A)\Big(\frac{27\alpha_{1}}{20}\nonumber\\
&&\hspace*{-5mm} + \frac{9\alpha_{2}}{4}\Big)\bigg]
- \frac{1}{(16\pi^2)^2}\bigg[\, 2\,\mbox{tr}\Big((Y_{U}Y_{U}^+)^2\Big)
+ 31\,\mbox{tr}\Big((Y_{D}Y_{D}^+)^2\Big) + 9\,\mbox{tr}\Big((Y_{E}Y_{E}^+)^2\Big)\nonumber\\
&&\hspace*{-5mm} +11\,\mbox{tr}\Big(Y_{D} Y_{D}^+ Y_{U} Y_{U}^+\Big)
+ 3 \Big(\mbox{tr}(Y_{U} Y_{U}^+)\Big)^2
+ 9 \Big(\mbox{tr}(Y_{D} Y_{D}^+)\Big)^2
+ 3\,\mbox{tr}(Y_{E} Y_{E}^+)\,\mbox{tr}(Y_{D} Y_{D}^+) \bigg]\nonumber\\
&&\hspace*{-5mm} + O\Big(\alpha^3,\alpha^2 Y^2,\alpha Y^4, Y^6\Big);\\
&&\vphantom{1}\nonumber\\
\label{Gamma_DetYE_HD+MSL}
&&\hspace*{-5mm} \gamma_{\mbox{\scriptsize det}\,Y_E}(\alpha,Y) = \frac{1}{2}\Big(3\gamma_{H_d}(\alpha,Y)  + \mbox{tr}\,\gamma_{L}(\alpha,Y) + \mbox{tr}\,\gamma_E(\alpha,Y) \Big)\nonumber\\
&&\hspace*{-5mm} = -\frac{27\alpha_1}{20\pi} - \frac{9\alpha_2}{4\pi}
+ \frac{1}{16\pi^2}\, \mbox{tr}\Big( 9\,Y_D Y_D^+ + 6\,Y_E Y_E^+\Big)
+ \frac{1}{2\pi^2}\bigg[\frac{243\alpha_1^2}{400}
+ \frac{27\alpha_2^2}{16} + \frac{27\alpha_1\alpha_2}{40}
\nonumber\\
&&\hspace*{-5mm}
-\frac{27\alpha_2^2}{2}\Big(\ln a_{\varphi,2}+1+\frac{A}{2}\Big) + \frac{891\alpha_1^2}{100}\Big(\ln a_{1}+1+\frac{A}{2}\Big) + \frac{63\alpha_2^2}{4}\Big(\ln a_{2}+1+\frac{A}{2}\Big) \bigg] + \frac{1}{16\pi^3}\nonumber\\
&&\hspace*{-5mm} \times\, \mbox{tr}(Y_{D} Y_{D}^+) \bigg[-\frac{3\alpha_{1}}{10} + 12\alpha_{3} + (B-A)\Big(\frac{21\alpha_{1}}{20} + \frac{27\alpha_{2}}{4} + 12\alpha_{3}\Big)\bigg]
+ \frac{1}{16\pi^3} \mbox{tr}\Big(Y_{E} Y_{E}^+\Big)\nonumber\\
&&\hspace*{-5mm}\times\,\bigg[\frac{9\alpha_{1}}{10} + \frac{3\alpha_2}{2} + (B-A)\Big(\frac{27\alpha_{1}}{10} + \frac{9\alpha_{2}}{2}\Big)\bigg]
- \frac{1}{(16\pi^2)^2}\bigg[\, 27\,\mbox{tr}\Big((Y_{D}Y_{D}^+)^2\Big)
+ 13\,\mbox{tr}\Big((Y_{E}Y_{E}^+)^2\Big)\nonumber\\
&&\hspace*{-5mm} +9\,\mbox{tr}\Big(Y_{D} Y_{D}^+ Y_{U} Y_{U}^+\Big)
+ 3 \Big(\mbox{tr}(Y_{E} Y_{E}^+)\Big)^2
+ 9\,\mbox{tr}(Y_{E} Y_{E}^+)\,\mbox{tr}(Y_{D} Y_{D}^+) \bigg] + O\Big(\alpha^3,\alpha^2 Y^2,\alpha Y^4, Y^6\Big);\nonumber\\
&& \vphantom{1}\\
\label{Gamma_Mu_HD+MSL}
&&\hspace*{-5mm} \gamma_{\bm{\mu}}(\alpha,Y) = \frac{1}{2}\Big(\gamma_{H_u}(\alpha,Y)  + \gamma_{H_d}(\alpha,Y) \Big)\nonumber\\
&&\hspace*{-5mm} = - \frac{3\alpha_{1}}{20\pi} - \frac{3\alpha_{2}}{4\pi} + \frac{1}{16\pi^2}\, \mbox{tr}\Big(3\, Y_{U} Y_{U}^+ + 3\, Y_{D} Y_{D}^+ + Y_{E} Y_{E}^+\Big)  + \frac{1}{2\pi^2}\bigg[ \frac{9\alpha_{1}^2}{400} + \frac{9\alpha_{2}^2}{16} + \frac{9}{40} \alpha_{1}\alpha_{2}\nonumber\\
&&\hspace*{-5mm} -\frac{9\alpha_{2}^2}{2}\Big(\ln a_{\varphi, 2} + 1+\frac{A}{2} \Big) + \frac{99\alpha_{1}^2}{100} \Big(\ln a_{1} + 1+\frac{A}{2} \Big)
+ \frac{21\alpha_{2}^2}{4} \Big(\ln a_{2} + 1 +\frac{A}{2} \Big)\bigg] + \frac{1}{16\pi^3}\nonumber\\
&&\hspace*{-5mm} \times\,\mbox{tr}\Big(Y_{E} Y_{E}^+\Big)\bigg[\frac{3\alpha_{1}}{10} + (B-A)\Big(\frac{9\alpha_{1}}{20} + \frac{3\alpha_{2}}{4}\Big)\bigg]
+\frac{1}{16\pi^3} \mbox{tr}\Big(Y_{U} Y_{U}^+\Big) \bigg[\frac{\alpha_{1}}{5} + 4\alpha_{3} +(B-A)\nonumber\\
&&\hspace*{-5mm} \times \Big(\frac{13\alpha_{1}}{20} + \frac{9\alpha_{2}}{4} + 4\alpha_{3}\Big)\bigg]
+ \frac{1}{16\pi^3} \mbox{tr}\Big(Y_{D} Y_{D}^+\Big)\bigg[-\frac{\alpha_{1}}{10} + 4\alpha_{3} + (B-A)\Big(\frac{7\alpha_{1}}{20} + \frac{9\alpha_{2}}{4} + 4\alpha_{3}\Big)\bigg] \nonumber\\
&&\hspace*{-5mm}  - \frac{1}{(16\pi^2)^2}\bigg[3\, \mbox{tr}\Big((Y_{E} Y_{E}^+)^2\Big) + 6\, \mbox{tr}\Big(Y_{D} Y_{D}^+ Y_{U} Y_{U}^+\Big) + 9\,\mbox{tr}\Big((Y_{U}Y_{U}^+)^2\Big) + 9\,\mbox{tr}\Big((Y_{D} Y_{D}^+)^2\Big)\bigg]
\nonumber\\
&&\hspace*{-5mm} + O\Big(\alpha^3,\alpha^2 Y^2,\alpha Y^4, Y^6\Big).\vphantom{\frac{1}{2}}
\end{eqnarray}

\section{Expressions for RGFs in the $\overline{\mbox{DR}}$ scheme}
\hspace*{\parindent}\label{Appendix_RGFs_DR}

To verify Eqs. (\ref{Third_Equation}) and (\ref{Fourth_Equation}) in the $\overline{\mbox{DR}}$ scheme, we present the expressions for the three-loop MSSM $\beta$-functions calculated in \cite{Jack:2004ch} and rewritten in the notation adopted in this paper,

\begin{eqnarray}\label{Beta1_DR}
&&\hspace*{-5mm} \frac{\beta_1(\alpha,Y)}{\alpha_{1}^2} = - \frac{1}{2\pi}\cdot \frac{3}{5} \bigg\{ -11 -\frac{199\alpha_{1}}{60\pi} -\frac{9\alpha_{2}}{4\pi} -\frac{22\alpha_{3}}{3\pi}
+ \frac{1}{8\pi^2} \mbox{tr}\Big(\frac{13}{3} Y_{U} Y_{U}^+ + \frac{7}{3} Y_{D} Y_{D}^+ + 3 Y_{E} Y_{E}^+ \Big) \nonumber\\
&&\hspace*{-5mm}
+ \frac{1}{2\pi^2}\Big( \frac{32117\alpha_{1}^2}{1800} + \frac{27\alpha_{2}^2}{8} - \frac{121\alpha_{3}^2}{18} + \frac{23\alpha_{1}\alpha_{2}}{40} + \frac{137\alpha_{1}\alpha_{3}}{45} + \alpha_{2}\alpha_{3}\Big)
+ \frac{1}{8\pi^3}\mbox{tr}\Big(Y_{U} Y_{U}^+\Big)\Big(\frac{169\alpha_1}{360}\nonumber\\
&&\hspace*{-5mm} + \frac{29\alpha_{2}}{8} + \frac{44\alpha_{3}}{9}\Big)
+ \frac{1}{8\pi^3}\mbox{tr}\Big(Y_{D} Y_{D}^+\Big)\Big( \frac{49\alpha_1}{360} + \frac{11\alpha_{2}}{8} + \frac{32\alpha_{3}}{9}\Big)
+ \frac{1}{8\pi^3}\mbox{tr}\Big(Y_{E} Y_{E}^+\Big)\Big(\frac{27\alpha_1}{40} + \frac{21\alpha_{2}}{8} \Big)
\nonumber\\
&&\hspace*{-5mm}
- \frac{1}{(8\pi^2)^2}\bigg[\,7\,\mbox{tr}\Big((Y_{U} Y_{U}^+)^2\Big) + \frac{9}{2} \mbox{tr}\Big((Y_{D} Y_{D}^+)^2\Big) + \frac{9}{2} \mbox{tr}\Big((Y_{E} Y_{E}^+)^2\Big) + \frac{29}{6}\mbox{tr}\Big(Y_{D} Y_{D}^+ Y_{U} Y_{U}^+\Big)   \nonumber\\
&&\hspace*{-5mm} + \frac{15}{2} \Big(\mbox{tr}(Y_{U} Y_{U}^+)\Big)^2 + 3 \Big(\mbox{tr}(Y_{D} Y_{D}^+)\Big)^2 + 2 \Big(\mbox{tr}(Y_{E} Y_{E}^+)\Big)^2
+ 7\, \mbox{tr}\Big(Y_{E} Y_{E}^+\Big)\,\mbox{tr}\Big(Y_{D} Y_{D}^+\Big) \bigg]\bigg\}\nonumber\\
&&\hspace*{-5mm} + O(\alpha^3,\alpha^2 Y^2, \alpha Y^4, Y^6);\nonumber\\
&&\vphantom{1}\\
\label{Beta2_DR}
&&\hspace*{-5mm} \frac{\beta_2(\alpha,Y)}{\alpha_{2}^2} = - \frac{1}{2\pi} \bigg\{-1  -\frac{9\alpha_{1}}{20\pi}- \frac{25\alpha_{2}}{4\pi} -\frac{6\alpha_{3}}{\pi} + \frac{1}{8\pi^2}\,
\mbox{tr}\Big(3\, Y_{U} Y_{U}^+ + 3\, Y_{D} Y_{D}^+ + Y_{E} Y_{E}^+ \Big) + \frac{1}{2\pi^2}\nonumber\\
&&\hspace*{-5mm} \times\Big( \frac{457\alpha_{1}^2}{200} - \frac{35\alpha_{2}^2}{8} - \frac{11\alpha_{3}^2}{2} - \frac{9\alpha_{1}\alpha_{2}}{40} + \frac{\alpha_{1}\alpha_{3}}{5} - 3\alpha_{2}\alpha_{3}\Big)
+ \frac{1}{8\pi^3}\mbox{tr}\Big(Y_{U} Y_{U}^+\Big)\Big( \frac{29\alpha_{1}}{40} + \frac{33\alpha_{2}}{8}  \nonumber\\
&&\hspace*{-5mm}
+ 4\alpha_{3}\Big) + \frac{1}{8\pi^3}\mbox{tr}\Big(Y_{D} Y_{D}^+\Big)\Big( \frac{11\alpha_{1}}{40}+\frac{33\alpha_{2}}{8} +4\alpha_{3}\Big)
+ \frac{1}{8\pi^3}\mbox{tr}\Big(Y_{E} Y_{E}^+\Big)\Big(\frac{21\alpha_{1}}{40} + \frac{11\alpha_{2}}{8}\Big)
- \frac{1}{(8\pi^2)^2}\nonumber\\
&&\hspace*{-5mm} \times \bigg[\, 6\,\mbox{tr}\Big((Y_{U} Y_{U}^+)^2\Big) + 6\,\mbox{tr}\Big((Y_{D} Y_{D}^+)^2\Big)
+ 2\,\mbox{tr}\Big((Y_{E} Y_{E}^+)^2\Big) + \frac{9}{2}\Big(\mbox{tr}(Y_{U} Y_{U}^+)\Big)^2 + \frac{9}{2}\Big(\mbox{tr}(Y_{D} Y_{D}^+)\Big)^2
\nonumber\\
&&\hspace*{-5mm}
+ \frac{1}{2}\Big(\mbox{tr}(Y_{E} Y_{E}^+)\Big)^2  + 3\, \mbox{tr}\Big(Y_{E} Y_{E}^+\Big)\, \mbox{tr}\Big(Y_{D} Y_{D}^+\Big)
+ 3\,\mbox{tr}\Big(Y_{D} Y_{D}^+ Y_{U} Y_{U}^+\Big) \bigg\} + O\Big(\alpha^3,\alpha^2 Y^2, \alpha Y^4, Y^6\Big);\nonumber\\
&&\vphantom{1}\\
\label{Beta3_DR}
&&\hspace*{-5mm} \frac{\beta_3(\alpha,Y)}{\alpha_{3}^2} = - \frac{1}{2\pi} \bigg\{3 -\frac{11\alpha_{1}}{20\pi} -\frac{9\alpha_{2}}{4\pi} -\frac{7\alpha_{3}}{2\pi}
+ \frac{1}{8\pi^2}\, \mbox{tr}\Big(2\, Y_{U} Y_{U}^+ + 2\, Y_{D} Y_{D}^+\Big) + \frac{1}{2\pi^2}\Big( \frac{851\alpha_{1}^2}{300}\nonumber\\
&&\hspace*{-5mm}  + \frac{27\alpha_{2}^2}{8} - \frac{347\alpha_{3}^2}{24} + \frac{3\alpha_{1}\alpha_{2}}{40} - \frac{11\alpha_{1}\alpha_{3}}{60}
-\frac{3\alpha_{2}\alpha_{3}}{4} \Big) + \frac{1}{8\pi^3}\mbox{tr}\Big(Y_{U} Y_{U}^+\Big)\Big(\frac{11\alpha_{1}}{30} + \frac{3\alpha_{2}}{2} + \frac{13\alpha_{3}}{3} \Big)\nonumber\\
&&\hspace*{-5mm} + \frac{1}{8\pi^3}\mbox{tr}\Big(Y_{D} Y_{D}^+\Big)\Big( \frac{4\alpha_{1}}{15} + \frac{3\alpha_{2}}{2}
+ \frac{13\alpha_{3}}{3}\Big) - \frac{1}{(8\pi^2)^2}\bigg[\, 3\, \mbox{tr}\Big((Y_{U} Y_{U}^+)^2\Big) + 3\, \mbox{tr}\Big((Y_{D} Y_{D}^+)^2\Big)\nonumber\\
&&\hspace*{-5mm} + \frac{9}{2} \Big(\mbox{tr}(Y_{U} Y_{U}^+)\Big)^2
+ \frac{9}{2}\Big(\mbox{tr}(Y_{D} Y_{D}^+)\Big)^2 + \frac{3}{2}\,\mbox{tr}\Big(Y_{E} Y_{E}^+\Big)\,\mbox{tr}\Big(Y_{D} Y_{D}^+\Big) + 2\,\mbox{tr}\Big(Y_{D} Y_{D}^+ Y_{U} Y_{U}^+\Big) \bigg]\bigg\}
\nonumber\\
&&\hspace*{-5mm} + O\Big(\alpha^3,\alpha^2 Y^2, \alpha Y^4, Y^6\Big).
\end{eqnarray}

The anomalous dimensions entering Eqs. (\ref{Third_Equation}) and (\ref{Fourth_Equation}) can also be calculated with the help of Eqs. (\ref{Determinant_Equations}) and (\ref{Mu_Renormalization}) because in the $\overline{\mbox{DR}}$ scheme all renormalization constants contain only $\varepsilon$-poles. The two-loop anomalous dimensions for all MSSM chiral matter superfields were first calculated in \cite{Bjorkman:1985mi}. They
are also given in \cite{Jack:2004ch}. In the notation adopted in this paper the corresponding results can be found in \cite{Haneychuk:2022qvu}. Using them, after some algebraic transformations we obtain the anomalous dimensions needed for checking Eqs. (\ref{Third_Equation}) and (\ref{Fourth_Equation}) for the $\overline{\mbox{DR}}$ renormalization prescription,

\begin{eqnarray}\label{Gamma_DetYU_DR}
&&\hspace*{-5mm} \gamma_{\mbox{\scriptsize det}\,Y_U}(\alpha,Y) = \frac{1}{2}\Big(3\gamma_{H_u}(\alpha,Y)  + \mbox{tr}\,\gamma_{Q}(\alpha,Y) + \mbox{tr}\,\gamma_U(\alpha,Y) \Big)\nonumber\\
&&\hspace*{-5mm} = -\frac{13\alpha_1}{20\pi} - \frac{9\alpha_2}{4\pi} - \frac{4\alpha_3}{\pi} + \frac{1}{16\pi^2}\, \mbox{tr}\Big(12\, Y_U Y_U^+ + Y_D Y_D^+\Big) + \frac{1}{2\pi^2}\bigg[\frac{2743\alpha_1^2}{1200}
+ \frac{45\alpha_2^2}{16}- \frac{2\alpha_3^2}{3}\nonumber\\
&&\hspace*{-5mm} + \frac{3\alpha_1\alpha_2}{8} + \frac{17\alpha_1\alpha_3}{15} + 3\alpha_2\alpha_3  \bigg] + \frac{1}{16\pi^3}\, \mbox{tr}(Y_{U} Y_{U}^+) \bigg[\frac{7\alpha_1}{10} +\frac{3\alpha_2}{2}+12\alpha_3 \bigg]
+ \frac{1}{16\pi^3} \mbox{tr}(Y_{D} Y_{D}^+)\nonumber\\
&&\hspace*{-5mm} \times\,\frac{\alpha_{1}}{10} - \frac{1}{(16\pi^2)^2}\bigg[\, 31\,\mbox{tr}\Big((Y_{U}Y_{U}^+)^2\Big) + 2\,\mbox{tr}\Big((Y_{D}Y_{D}^+)^2\Big)
+11\,\mbox{tr}\Big(Y_{D} Y_{D}^+ Y_{U} Y_{U}^+\Big) + 9 \Big(\mbox{tr}(Y_{U} Y_{U}^+)\Big)^2\nonumber\\
&&\hspace*{-5mm} + 3\Big(\mbox{tr}(Y_{D} Y_{D}^+)\Big)^2 + \mbox{tr}(Y_{E} Y_{E}^+)\,\mbox{tr}(Y_{D} Y_{D}^+)\bigg] + O\Big(\alpha^3,\alpha^2 Y^2,\alpha Y^4, Y^6\Big);\\
&&\vphantom{1}\nonumber\\
\label{Gamma_DetYD_DR}
&&\hspace*{-5mm} \gamma_{\mbox{\scriptsize det}\,Y_D}(\alpha,Y) = \frac{1}{2}\Big(3\gamma_{H_d}(\alpha,Y)  + \mbox{tr}\,\gamma_{Q}(\alpha,Y) + \mbox{tr}\,\gamma_D(\alpha,Y) \Big)\nonumber\\
&&\hspace*{-5mm} = -\frac{7\alpha_1}{20\pi} - \frac{9\alpha_2}{4\pi} - \frac{4\alpha_3}{\pi}
+ \frac{1}{16\pi^2}\, \mbox{tr}\Big( Y_U Y_U^+ + 12\,Y_D Y_D^+ + 3\,Y_E Y_E^+\Big) + \frac{1}{2\pi^2}\bigg[\frac{287\alpha_1^2}{240}
+ \frac{45\alpha_2^2}{16} \nonumber\\
&&\hspace*{-5mm} - \frac{2\alpha_3^2}{3} + \frac{3\alpha_1\alpha_2}{8}
+ \frac{\alpha_1\alpha_3}{3} + 3\alpha_2\alpha_3\bigg]
+ \frac{1}{16\pi^3} \mbox{tr}(Y_{U} Y_{U}^+) \cdot\frac{\alpha_{1}}{5}
+ \frac{1}{16\pi^3} \mbox{tr}(Y_{D} Y_{D}^+) \bigg[-\frac{\alpha_{1}}{10} + \frac{3\alpha_2}{2}
\nonumber\\
&&\hspace*{-5mm}
+ 12\alpha_3 \bigg] + \frac{1}{16\pi^3} \mbox{tr}\Big(Y_{E} Y_{E}^+\Big)\cdot\frac{9\alpha_{1}}{10}
- \frac{1}{(16\pi^2)^2}\bigg[\, 2\,\mbox{tr}\Big((Y_{U}Y_{U}^+)^2\Big)
+ 31\,\mbox{tr}\Big((Y_{D}Y_{D}^+)^2\Big) \nonumber\\
&&\hspace*{-5mm} + 9\,\mbox{tr}\Big((Y_{E}Y_{E}^+)^2\Big)
+11\,\mbox{tr}\Big(Y_{D} Y_{D}^+ Y_{U} Y_{U}^+\Big)
+ 3 \Big(\mbox{tr}(Y_{U} Y_{U}^+)\Big)^2
+ 9 \Big(\mbox{tr}(Y_{D} Y_{D}^+)\Big)^2
+ 3\,\mbox{tr}(Y_{E} Y_{E}^+)\nonumber\\
&&\hspace*{-5mm}\times\,\mbox{tr}(Y_{D} Y_{D}^+) \bigg] + O\Big(\alpha^3,\alpha^2 Y^2,\alpha Y^4, Y^6\Big);\\
&&\vphantom{1}\nonumber\\
\label{Gamma_DetYE_DR}
&&\hspace*{-5mm} \gamma_{\mbox{\scriptsize det}\,Y_E}(\alpha,Y) = \frac{1}{2}\Big(3\gamma_{H_d}(\alpha,Y)  + \mbox{tr}\,\gamma_{L}(\alpha,Y) + \mbox{tr}\,\gamma_E(\alpha,Y) \Big)\nonumber\\
&&\hspace*{-5mm} = -\frac{27\alpha_1}{20\pi} - \frac{9\alpha_2}{4\pi}
+ \frac{1}{16\pi^2}\, \mbox{tr}\Big( 9\,Y_D Y_D^+ + 6\,Y_E Y_E^+\Big)
+ \frac{1}{2\pi^2}\bigg[\frac{81\alpha_1^2}{16} + \frac{45\alpha_2^2}{16} + \frac{27\alpha_1\alpha_2}{40} \bigg]
+ \frac{1}{16\pi^3}\nonumber\\
&&\hspace*{-5mm} \times\, \mbox{tr}(Y_{D} Y_{D}^+) \bigg[-\frac{3\alpha_{1}}{10} + 12\alpha_{3} \bigg]
+ \frac{1}{16\pi^3} \mbox{tr}\Big(Y_{E} Y_{E}^+\Big) \bigg[\frac{9\alpha_{1}}{10} + \frac{3\alpha_2}{2} \bigg]
- \frac{1}{(16\pi^2)^2}\bigg[\, 27\,\mbox{tr}\Big((Y_{D}Y_{D}^+)^2\Big)\nonumber\\
&&\hspace*{-5mm}
+ 13\,\mbox{tr}\Big((Y_{E}Y_{E}^+)^2\Big)
+9\,\mbox{tr}\Big(Y_{D} Y_{D}^+ Y_{U} Y_{U}^+\Big)
+ 3 \Big(\mbox{tr}(Y_{E} Y_{E}^+)\Big)^2
+ 9\,\mbox{tr}(Y_{E} Y_{E}^+)\,\mbox{tr}(Y_{D} Y_{D}^+) \bigg]\nonumber\\
&&\hspace*{-5mm} + O\Big(\alpha^3,\alpha^2 Y^2,\alpha Y^4, Y^6\Big);\\
&& \vphantom{1}\nonumber\\
\label{Gamma_Mu_DR}
&&\hspace*{-5mm} \gamma_{\bm{\mu}}(\alpha,Y) = \frac{1}{2}\Big(\gamma_{H_u}(\alpha,Y)  + \gamma_{H_d}(\alpha,Y) \Big)\nonumber\\
&&\hspace*{-5mm} = - \frac{3\alpha_{1}}{20\pi} - \frac{3\alpha_{2}}{4\pi} + \frac{1}{16\pi^2}\, \mbox{tr}\Big(3\, Y_{U} Y_{U}^+ + 3\, Y_{D} Y_{D}^+ + Y_{E} Y_{E}^+\Big)  + \frac{1}{2\pi^2}\Big(\frac{207\alpha_{1}^2}{400} + \frac{15\alpha_{2}^2}{16} + \frac{9\alpha_{1}\alpha_{2}}{40}\Big) \nonumber\\
&&\hspace*{-5mm} + \frac{1}{16\pi^3}\,\bigg[\,\mbox{tr}\Big(Y_{E} Y_{E}^+\Big) \frac{3\alpha_{1}}{10}
+ \mbox{tr}\Big(Y_{U} Y_{U}^+\Big) \Big(\frac{\alpha_{1}}{5} + 4\alpha_{3} \Big)
+ \mbox{tr}\Big(Y_{D} Y_{D}^+\Big)\Big(-\frac{\alpha_{1}}{10} + 4\alpha_{3}\Big)\bigg] \nonumber\\
&&\hspace*{-5mm}  - \frac{1}{(16\pi^2)^2}\bigg[3\, \mbox{tr}\Big((Y_{E} Y_{E}^+)^2\Big) + 6\, \mbox{tr}\Big(Y_{D} Y_{D}^+ Y_{U} Y_{U}^+\Big) + 9\,\mbox{tr}\Big((Y_{U}Y_{U}^+)^2\Big) + 9\,\mbox{tr}\Big((Y_{D} Y_{D}^+)^2\Big)\bigg]
\nonumber\\
&&\hspace*{-5mm} + O\Big(\alpha^3,\alpha^2 Y^2,\alpha Y^4, Y^6\Big).\vphantom{\frac{1}{2}}
\end{eqnarray}

\end{document}